\documentclass[12pt]{article}
\usepackage[english]{babel}
\usepackage{cite}
\usepackage{amsmath,amssymb}
\usepackage{epsfig}

\usepackage{tikz}
\usetikzlibrary{matrix,positioning,decorations.pathreplacing}

\renewcommand{\d}{\mathrm{d}}
\begin{document}
\bibliographystyle{unsrt}
	
\thispagestyle{empty}

\vspace*{2.0cm}

\begin{center}
  \vspace{-3cm}{\baselineskip14pt                                                    \centerline{\normalsize DESY 22-122\hfill ISSN 0418-9833}                       \centerline{\normalsize July 2022\hfill}}
\vspace{1cm} 
  \boldmath
{\large \bf
Specializations of partial differential equations for Feynman integrals}
\unboldmath
\end{center}
 \vspace*{0.8cm}

\begin{center}
{\sc Vladimir~V.~Bytev$^{b}$\footnote{E-mail: bvv@jinr.ru}
Bernd~A.~Kniehl$^a$\footnote{E-mail: kniehl@desy.de}
Oleg~L.~Veretin$^a$\footnote{E-mail: veretin@mail.desy.de}}

\vspace*{1.0cm}

{\normalsize $^{a}$ II. Institut f\"ur Theoretische Physik, Universit\"at Hamburg,}\\
{\normalsize Luruper Chaussee 149, 22761 Hamburg, Germany} \\
\bigskip
{\normalsize $^{b}$ Joint Institute for Nuclear Research,} \\
{\normalsize $141980$ Dubna (Moscow Region), Russia}
\end{center}

\begin{abstract}
Starting from the Mellin--Barnes integral representation of a Feynman integral
depending on a set of kinematic variables $z_i$, we derive a system of partial
differential equations w.r.t.\ new variables $x_j$, which parameterize the
differentiable constraints $z_i=y_i(x_j)$.
In our algorithm, the powers of propagators can be considered as arbitrary
parameters.
Our algorithm can also be used for the reduction of multiple hypergeometric
sums to sums of lower dimension, finding special values and reduction
equations of hypergeometric functions in a singular locus of continuous
variables, or finding systems of partial differential equations for master
integrals with arbitrary powers of propagators.
As an illustration, we produce a differential equation of fourth order in one variable for the one-loop two-point Feynman diagram with two different masses and arbitrary propagator powers.
\end{abstract}

\newpage

\section{Introduction}

Within dimensional regularization, the general $L$-loop Feynman integral (FI)
with $N$ internal momenta $q_i$ and masses $m_i$, and $E$ external momenta
$p_i$ is given by
\begin{align}
J( \{s\}, \{m\}, \{\alpha\}) =
  \int\frac{{\rm d}^d k_1\dots {\rm d}^d k_L}{(i\pi^{d/2})^L} \prod\limits_{i=1}^N\frac{1}{(q_i^2-m_i^2)^{\alpha_i}}\,,
\label{eq:defJ}
\end{align}
where $\{s\}$ is the full  set of external Lorentz invariants constructed from the
external momenta,
\begin{align}
  s_{ij} = p_i\cdot p_j\,, \qquad  i,j=1,\dots,E\,, \qquad i\le j \,,
\label{eq:invars}
\end{align}
and $\{m\}$, $\{\alpha\}$ are the sets of the $N$ masses $m_i$ and the $N$
indices $\alpha_i$, respectively.
The internal momenta $q_i$ are linear combinations of the loop momenta $k_j$
and external momenta $p_k$.
The indices $\alpha_i$ of the propagators are usually assumed to be integer
numbers, both positive and negative.
In Section~\ref{sec:two}, we will allow for $\alpha_i$ to be real numbers.

One of the most powerful approaches to the evaluation of FIs is
based on the method of differential equations
\cite{Kotikov:1990kg,Kotikov:1991hm,Kotikov:1991pm,Kotikov:1990zs,Remiddi:1997ny}.
Using the standard method of integration by parts (IBP) \cite{Chetyrkin:1981qh},
any FI (\ref{eq:defJ}) can be reduced to a set of so-called
master integrals with fixed sets of indices $\alpha_i$.
If we have $m$ master integrals, we can write a linear system of $m$ partial
differential equations (PDEs) of first order.
Alternatively, we can write a single linear differential equation of $m$-th
order.
In both cases, the $m$ solutions together with $m$ appropriate boundary
conditions provide us with representations of all master integrals. 

The conventional way to obtain the system of differential equations is to use
IBP reduction.
In fact, one can differentiate the master integrals w.r.t.\ some parameters
(masses or external Lorentz invariants) and then reduce these derivatives down
to the master integrals themselves.
In order for this program to work, one has to provide for each integral an
injective of $J$ into a special ``complete'' topology,\footnote{%
In the literature, also the terms ``auxiliary'' and ``full'' are used.}
which is minimal and complete in the following sense.
Any scalar product of any of the $L$ loop momenta with any other loop momentum
or any of the $E$ external momenta is expressible as a linear combination of
the propagators.
Such a complete topology has exactly $L(L+1)/2+LE$ propagators, which are
linearly independent.
Then, the particular FIs (\ref{eq:defJ}), usually called ``sector
integrals'', appear as special cases of the complete topology, where some of
the propagator indices become non-positive.

In this paper, we pursue the idea that the system of PDEs can be derived from
Mellin--Barnes representations without resorting to IBP relations.
In this case, we also do not need to construct the complete topologies, but can
obtain the differential equations directly in a given sector.
Moreover, usually all indices $\alpha_i$ in Eq.~(\ref{eq:defJ}) are taken to be
integers, including negative values, which account for possible numerators of
the FIs.
In this paper, however, we relax this restriction and consider $\alpha_i$ as
real numbers.

Applying the methods described, e.g., in
Refs.~\cite{Boos:1990rg,Smirnov:2004ym}, the expression in Eq.~(\ref{eq:defJ})
can be written as a multiple Mellin--Barnes representation in the form
\begin{align}
J( \{s\}, \{m\}, \{\alpha\}) &=  
      C \int\limits_{-i\infty}^{+i\infty}
	\prod\limits_{j,l} \,
	{\rm d}u_l \, \frac{\Gamma(\sum_{i}a_{ij} u_i+b_j)}{\Gamma(\sum_{i}c_{ij} u_i+d_j)}
	z_l^{\sum_k f_{kl}u_k},
	\label{eq:MB}
\end{align}
where $C$ is some constant, which depends on the propagator indices $\alpha_i$
and the space-time dimension $d$ \cite{tHooft:1972tcz}, and $z_j$ are ratios of
the external kinematic invariants in Eq.~(\ref{eq:invars}) and the masses
$m_i$.
All other parameters, $a_{ij}$, $b_{i}$, $c_{ij}$, $d_{i}$, and $f_{kl}$, are
linear combinations of the space-time dimension $d$ and the propagator indices
$\alpha_i$.
As usual, the integration is performed over the Mellin--Barnes parameters $u_k$ along the contours that separate the left and right poles in the complex $u_k$
planes.

The representation (\ref{eq:MB}) is our starting point.
In Refs.~\cite{Kalmykov:2008ofy,Kalmykov:2012rr}, it was already noted that,
using Eq.~(\ref{eq:MB}), we can obtain a system of PDEs in the variables $z_j$.
However, this approach is very restrictive in its practical applications.
In fact, it requires keeping all $z_j$ as independent variables, which is
usually not the case in real applications.
Typically, we have situations where there are either some relations between
these variables or some of them are fixed numbers and not subject to
differentiation
(for example, in the case of a single-scale diagram, 
or in the case where all masses are
equal).

In this paper, we show how, starting from representation~(\ref{eq:MB}), we can
obtain systems of PDEs with some constraints, $f_k(z_j)=0$
with $k=1,\dots,h$, which will allow us to consider cases of practical
interest.


\section{Algorithm}
\label{sec:two}

In this section, we describe in detail our algorithm for obtaining systems of
PDEs from the Mellin--Barnes representations (\ref{eq:MB}).

First, taking the residues of the $\Gamma(-u_l)$ functions
 at negative integer points, the Mellin--Barnes
integral (\ref{eq:MB}) can be written as a linear combination of
Horn-type hypergeometric series \cite{Kalmykov:2020cqz},
\begin{gather}
H({\bold a},\vec{b},{\bold c},\vec{d}; \,\vec{z})=\sum_{\vec{l}}	
	\prod\limits_j C_j
\frac{\Gamma(\sum_{i}a_{ij} l_i+b_j)}{\Gamma(\sum_{i}c_{ij} l_i+d_j)}\,
\frac{z_1^{l_1}}{l_1!}...\frac{z_n^{l_n}}{l_n!} \,,
\label{eq:2:2}
\end{gather}
where $C_j$ are some constants and we use the notations
${\bold a}=(a_{ij})$, ${\bold c}=(c_{ij})$, $\vec{b}=(b_j)$, $\vec{d}=(d_j)$,
etc. 

Shifting the integration contours in the Mellin--Barnes representation
(\ref{eq:MB}) or, equivalently, shifting the summation indices in the
hypergeometric representation (\ref{eq:2:2}), we obtain
{\it differential contiguous relations}.
These can be expressed in terms of step-up operators $L^+_{b_j}$ and step-down
operators $L^-_{d_j}$
\cite{Bytev:2009kb,Bytev:2011ks,Bytev:2013gva,Bytev:2013bqa,Bytev:2016ibi},
which shift the indices $b_j$, $d_j$ by one unit.
Specifically, we have
\begin{gather}
	H({\bold a},\vec{b}+\vec{e}_j,{\bold c},\vec{d}; \vec{z})
	= L^+_{b_j}H({\bold a},\vec{b},{\bold c},\vec{d}; \vec{z})
        \propto \Big( \sum_i a_{ij} \theta_i+b_j \Big)H({\bold a},\vec{b},{\bold c},\vec{d}; \,\vec{z})\,,
        \nonumber
        \\
	H ({\bold a},\vec{b},{\bold c},\vec{d}-\vec{e}_j; \vec{z})
	=L^-_{d_j}H({\bold a},\vec{b},{\bold c},\vec{d}; \,\vec{z})
        \propto \Big(\sum_i c_{ij} \theta_i+d_j-1 \Big)H({\bold a},\vec{b},{\bold c},\vec{d}; \vec{z})\,,
\label{eq:Lm}		
\end{gather}
where $\theta_i=z_i({\rm d}/{\rm d} z_i)$ is the Euler differential operator and
$\vec{e}_j$ are orthonormal unit basis vectors,
with $\vec{e}_j\cdot\vec{e}_k=\delta_{jk}$.

From the differential contiguous relations (\ref{eq:Lm}), a dynamical symmetry
algebra may be constructed.
From this Lie algebra, a system of PDEs that is satisfied by the function in
Eq.~(\ref{eq:2:2}) may be constructed \cite{Kalnins1980TheLT}.
More precisely, combining the operators $L^+_{b_j}$ and $L^-_{d_j}$ from
Eq.~(\ref{eq:Lm}) and differentiating Eq.~(\ref{eq:2:2}) w.r.t.\ variables
$z_j$, we can derive a system of $n$ PDEs:
\begin{gather}
  \left(\underset{j\in m^+_k,n_j\in\{0,a_{kj}\}}\prod L^+_{b_j+n_j}-\frac{1}{z_k}\theta_k
      \underset{j\in m^-_k,  n_j\in\{0,c_{kj}-1\}   }\prod L^-_{d_j-n_j        } \right)
      H({\bold a},\vec{b},{\bold c},\vec{d}; \vec{z})=0\,,
\label{main:diff:syst}
\end{gather}
where set $m^+_k$ consists of the integers $j$ for which
$a_{kj}\ne 0$ and set $m^-_k$
consists of the integers $j$ for which $c_{kj}\ne0$.
We imply here that variables $a_{ij}$ and $c_{ij}$ are natural numbers.

This system of PDEs may be derived directly from the Mellin--Barnes
representation (\ref{eq:MB}) \cite{Kalmykov:2012rr}.
In some cases, however, to obtain the full system of PDEs, a prolongation
procedure has to be applied, which consists of applying additional derivatives
to the system of PDEs to find one or more new nontrivial equations.
 
In the system of PDEs (\ref{main:diff:syst}), it is implied that the
variables $z_j$ are independent, i.e.\ all masses and external Lorentz
invariants are different and not equal to zero, and that the propagator
indices are real numbers. 

As for multivariate specializations of the PDE system (\ref{main:diff:syst}),
we have to consider several different cases \cite{VIDUNAS2009145}.
In the first case, the multivariate specialization falls into a singular locus
of the PDE system (\ref{main:diff:syst}).
Then, the rank of the new PDE system will be lower than the initial one, for
any combination of the parameters ${\bold a},\vec{b},{\bold c},\vec{d}$.

The singular loci of the new PDEs are inherited from the old ones with the old variables $z_i$  and induced locus of multivariate specialization.
For some particular combinations of parameters and variables, the loci of the
new PDEs could be diminished.

All this may be directly inferred from the PDE system (\ref{main:diff:syst}),
and the final differential equation(s) is/are satisfied by some hypergeometric
functions of lower (simpler) class. 
 
For any other multivariate specialization, the rank of the new PDE system is
the same as that of the initial one.
Nonetheless, we may observe simplifications in the class of functions that
satisfy the new PDE system after the application of projective or more general
pull-back transformations of variables. 
 
Finally, in the case when the monodromy group of the PDE system is reduced,
which manifests itself in a factorization of the PDEs in
Eq.~(\ref{main:diff:syst}) or in new differential equation(s) after
multivariate specialization, some of the solutions reduce to rational ones,
and the remaining solutions may be expressed through hypergeometric functions
of lower order.

As already mentioned above, the variables $z_j$ with $j=1,\dots,n$ are all
independent in the PDE system (\ref{main:diff:syst}) and vary in some
differential manifold in $\mathbb{C}^n$.
Now suppose that we impose $r$ differentiable constraints and parametrize the
new manifold in terms of new independent variables $x_i$ with $i=1,\dots,k$,
that is
\begin{align}
  z_j &= y_j(\vec{x}), \qquad j=1,\dots,n  \nonumber\\
  \vec{x} &= (x_1,\dots, x_k), \qquad k<n \,. 
\end{align}
Our goal is to derive from the system (\ref{main:diff:syst}) of PDEs w.r.t.\
the variables $z_1,\dots,z_n$ a new system of PDEs w.r.t.\ the new variables
$x_i,\dots,x_k$.
In the following, we omit for brevity the indexed arguments of $H$ in
Eq.~(\ref{eq:2:2}) and use the shorthand notations
$H(\vec{z})=H({\bold a},\vec{b},{\bold c},\vec{d};\vec{z})$
and $H(\vec{x})=H({\bold a},\vec{b},{\bold c},\vec{d};\vec{z}(\vec{x}))$.
First, we note that
 \begin{gather}
   \frac{dH(\vec{x})}{dx_j}=\sum_{i=1}^{n}
   \frac{\partial H(\vec{z})}{\partial z_i}\,\frac{\partial y_i}{\partial x_j}.
 	\label{chain_rule}
 \end{gather}

 The rank of the PDE system w.r.t.\ the new variables $\vec{x}$ must be the same
as that of the initial PDE system (\ref{main:diff:syst}), if the new variables
$y_j(\vec{x})$ do not fall into the singular locus of the PDE system. 
If this is not the case, then the total rank of the new PDE system is lower,
and we do not consider such degenerate cases here for simplicity.
Furthermore, the order $\chi$ of derivatives w.r.t.\ the new variables
must be higher than or equal to the order $\eta$ in Eq.~(\ref{main:diff:syst}).
We do not specify $\chi$ at this point, but consider it as an unknown parameter
$\chi>\eta$.

By applying the chain rule (\ref{chain_rule}) $\chi$ times, we construct a PDE
system in which various derivatives of order less than or equal to $\chi$
w.r.t.\ new variables $\vec{x}$ are expressed in terms of derivatives
w.r.t.\ old variables $\vec{z}$ and derivatives of $y_j(\vec{x})$ functions.
We solve this system and express some of the high-order derivatives w.r.t.\ old
variables $\vec{z}$ through a mixture of derivatives w.r.t.\ new $\vec{x}$
variables, old $\vec{z}$ variables, and derivatives of known $y_j(\vec{x})$
functions,
\begin{gather}
\left.\frac{\partial^iH(\vec{z})}{\partial z_{j_1}\dots \partial z_{j_i}}\right|_{i\le\chi}
  =\sum_{k<i}A_{j_1\dots j_k} \frac{\partial^k  H(\vec{z})}{\partial z_{j_1}\dots\partial z_{j_k}}
     +\sum_{l\le\chi}B_{j_1\dots j_l} 
\frac{\partial^l  H(\vec{x})}{\partial x_{j_1}\dots \partial x_{j_l}}\,.
	\label{subst_matrix}
\end{gather}

We now proceed with the derivation of Eq.~(\ref{main:diff:syst}) w.r.t.\
variables $\vec{z}$ through the combined order $\chi-\eta$.
Substituting derivatives w.r.t.\ old variables using Eq.~(\ref{subst_matrix}),
we construct the matrix $M$ of the PDE system, in which old and new variables
are mixed:
\begin{equation}
\begin{tikzpicture}[
	style1/.style={
		matrix of math nodes,
		every node/.append style={text width=#1,align=center,minimum height=5ex},
		nodes in empty cells,
		left delimiter=[,
		right delimiter=],
	},
	style2/.style={
		matrix of math nodes,
		every node/.append style={text width=#1,align=center,minimum height=5ex},
		nodes in empty cells,
		left delimiter=[,
		right delimiter=],
	}
	]
	\matrix[style1=0.85cm] (1mat)
	{
		& & & & &  \\
		& & & & & \\
		& & & & & \\
		& & & & & \\
		& & & & & \\
		& & & & & \\
		& & & & & \\
		& & & & & \\
		& & & & & \\
	};
	\draw[dashed]
	(1mat-6-1.south west) -- (1mat-6-6.south east);
	\draw[dashed]
	(1mat-1-4.north east) -- (1mat-9-4.south east);
	\node[font=\huge] 
	at (1mat-3-2.south east) {$M_{1}$};
	\node[font=\Large] 
	at (1mat-8-5.east) {$M_{2}$};
	\node[font=\Large] 
	at (1mat-8-2.east) {$B_{1}$};
	\node[font=\Large] 
	at (1mat-3-5.south east) {$*$};

	\matrix[style2=1.2cm,right=40pt of 1mat] (2mat)
	{
			\frac{\partial^\chi  H(\vec{z})}{\partial z_{j_1}...\partial z_{j_\chi}} \\
		\vdots \\
		\frac{\partial^i  H(\vec{z})}{\partial z_{j_1}...\partial z_{j_i}} \\
		\vdots \\
			\frac{\partial  H(\vec{z})}{\partial z_{j}} \\
\\
\frac{\partial^l  H(\vec{x})}{\partial x_{j_1}... \partial x_{j_l}} \\
	\vdots	\\
 H(\vec{x})	\\
	};
	\draw[dashed]
	(2mat-6-1.south west) -- (2mat-6-1.south east);
	
	\node at ([xshift=17pt,yshift=-1.2pt]2mat.east) {${}=0\,.$};

\end{tikzpicture}
\label{main_syst}
\end{equation}

If the rank of $M_1$ is less than that of $M$, then there exists a system of
independent equations that involves only derivatives w.r.t.\ new variables
$\vec{x}$.
If we now perform the row echelon reduction on $M$, the bottom-left block $B_1$
becomes zero.
Then, $M_2$ gives us an explicit form of the PDEs w.r.t.\ new variables
$\vec{x}$:
\begin{equation}
M_2\begin{bmatrix} 
\frac{\partial^l  H(\vec{x})}{\partial x_{j_1}\dots \partial x_{j_l}} \\
\vdots	\\
H(\vec{x})	\\
\end{bmatrix}
=0\,.
\end{equation}

However, if the rank of $M$ equals that of $M_1$,
then we have to increase the parameter $\chi$, i.e., the number of derivatives
w.r.t.\ new variables $\vec{x}$, and repeat the above procedure.

Let us now consider a special case, in which the constraints have the following
form:
\begin{gather}
  z_1=y(\vec{x})=x_1\,, \quad  z_2=y(\vec{x})=\mbox{const}_2\,,
    \quad \ldots \quad z_n=y_n(\vec{x})=\mbox{const}_n\,,
\end{gather}
i.e., we treat all variables $z_j$, except for the first one $z_1$, as
constants.
Then the maximum order $\chi$ of derivatives w.r.t.\ $x_1$ must be equal
to the differential rank of PDE system (\ref{main:diff:syst}).
Then the chain rule (\ref{chain_rule}) is trivial, and the PDE system
(\ref{subst_matrix}) contains $\chi$ different equations,
\begin{gather}
  \frac{\partial^i  H(\vec{z})}{\partial z_{j_1}\dots\partial  z_{j_i}}\biggr|_{i\le\chi}=
  \sum_{k<i}A_{j_1...j_k} \frac{\partial^k  H(\vec{z})}{\partial z_{j_1}\dots\partial z_{j_k}}+\sum_{l\le\chi}B_{j_1...j_l} 
\frac{\partial^l  H(\vec{x})}{\partial^l x_{1}}\,.
\label{eq:2:4}
\end{gather}
Applying $\chi-\eta$ differentiations w.r.t.\ variables $\vec{z}$ to
Eq.~(\ref{main:diff:syst}) and substituting $\chi$ derivatives from
Eq.~(\ref{eq:2:4}), we construct the matrix $M$ of PDEs in
Eq.~(\ref{main_syst}).
In the upper triangular form of the $M$ matrix, $B_1$ is zero and
$M_2=(Q_1,...,Q_\chi)^T$ is just a vector.
Thus, we arrive at an ordinary differential equation of order $\chi$
w.r.t.\ the single variable $x_1$,
\begin{gather}
	\sum_{i=0}^\chi Q_{\chi-i} \frac{\partial^i H(x_1) }{\partial^i x_1}=0\,.
\end{gather}

\section{Example}

We now illustrate our algorithm by means of a simple example.
Specifically, we derive the differential equation in one variable for the 
one-loop two-point Feynman diagram with different masses and arbitrary powers
of propagators,
\begin{gather}
  J(\alpha_1,\alpha_2,m_1,m_2)=
  \int\frac{{\rm d}^d k}{i\pi^{d/2}}\,\frac{1}{(k^2-m_1^2)^{\alpha_1}\left[(k-p)^2-m_2^2\right]^{\alpha_2}} \,.
\label{bubble_FI}
\end{gather}
We can rewrite each of the propagators as \cite{Boos:1990rg} 
\begin{gather}
\frac{1}{(k^2-m^2)^\beta}=\frac{1}{(k^2)^\beta}\frac{1}{\Gamma(\beta)}\frac{1}{2\pi i}\int_{-i\infty}^{+i\infty}
\d s \left(\frac{-m^2}{k^2}\right)^s \Gamma(-s)\Gamma(\beta+s) \,.
\end{gather}
Integrating over the massless one-loop propagator, we then obtain a 
two-fold Mellin--Barnes representation of Eq.~(\ref{bubble_FI}),
\begin{gather}
J(\alpha_1,\alpha_2,m_1,m_2)=\pi^{d/2}i^{1-d}(p^2)^{d/2-\alpha-\beta}\frac{1}{(2\pi i)^2}\,
\frac{1}{\Gamma(\alpha)\Gamma(\beta)}\int \d s \d u \left(\frac{p^2}{-m_1^2}\right)^s
\left(\frac{p^2}{-m_2^2}\right)^u
\nonumber \\
\times
\frac{\Gamma(d/2-\alpha+s)\Gamma(d/2-\beta+u)\Gamma(\alpha+\beta-d/2-s-u)\Gamma(s)\Gamma(u)}
{\Gamma(\alpha-s)\Gamma(\beta-u)\Gamma(d-\alpha-\beta+s+u)} \,.
\end{gather}

By constructing step-up and step-down operators according to Eq.~(\ref{eq:Lm})
and combining them with the differentiations w.r.t.\ $z_1=p^2/m_1^2$,
$z_2=p^2/m_2^2$, we obtain the following system of PDEs of second order in two
variables for $J(\alpha,\beta,m_1,m_2)$ \cite{Kalmykov:2012rr}:
\begin{gather}
 {\theta_1} \left(- {\alpha_1}+\frac{d}{2}+ {\theta_1}\right)-\frac{(2  {\alpha_1}+2  {\alpha_2}-d-2  {\theta_1}-2  {\theta_2}) ( {\alpha_1}+ {\alpha_2}-d- {\theta_1}- {\theta_2}+1)}{2 z_1}=0 \,,
 \nonumber \\
 {\theta_2} \left(- {\alpha_2}+\frac{d}{2}+ {\theta_2}\right)-\frac{(2  {\alpha_1}+2  {\alpha_2}-d-2  {\theta_1}-2  {\theta_2}) ( {\alpha_1}+ {\alpha_2}-d- {\theta_1}- {\theta_2}+1)}{2 z_2}=0 \,.
 \label{bubble:eq:th}
\end{gather}

Let us now construct an ordinary differential equation w.r.t.\ the variable
$z_1=x$.
To this end, we impose the constraints $z_1=y_1(x)=x$ and
$z_2=y_2(x)=\mbox{const}_2$.
Notice that the PDE system (\ref{bubble:eq:th}) is equivalent to the PDE system
\begin{gather}
\frac{1	}{z_1}{\theta_1} \left(c_1-1+ {\theta_1}\right)-(a+  {\theta_1}+  {\theta_2}) ( b+ {\theta_1}+ {\theta_2})=0\,,
	\nonumber \\
\frac{1}{z_2}	{\theta_2} \left( c_2-1+ {\theta_2}\right)-(a+  {\theta_1}+  {\theta_2}) (b+ {\theta_1}+ {\theta_2})=0
	\label{eq_F4}
\end{gather}
for the Appell hypergeometric function $F_4(a,b,c_1,c_2,z_1,z_2)$
\cite{Bytev:2016ibi},
with $a=\alpha_1+\alpha_2-d/2$, $b=1+\alpha_1+\alpha_2-d$,
$c_i=1+\alpha_i-d/2$, and has four different solutions.
The singular locus is $z_1=0$, $z_2=0$, the line at infinity, and
$z_1^2+z_2^2+1=2 z_1 z_2+2z_1+2z_2$, and, as the
  variables $y_1(x)$, $y_2(x)$ do not belong to the locus, the rank of the new
  PDE system is the same as that of the old one.
  Thus, we choose the number of derivatives w.r.t.\ $x$ to be $\chi=4$.
  In this case, we need $\chi-\eta=2$
  differentiations of the PDE system (\ref{bubble:eq:th}) w.r.t.\ variables
  $z_1,z_2$.
The algorithm of Section~\ref{sec:two} produces
the following differential equation of fourth order in $x$:
 \begin{gather}
 	L_4(x)J(\alpha_1,\alpha_2,m_1,m_2)=0 \,,
 	\label{eq:ex:4}
 \end{gather}
 where $L_4(x)$ is a differential operator of fourth order, whose expression
 is too lengthy to be listed here.

 As for Eq.~(\ref{eq_F4}), the region of the exceptional set of parameters,
 when the monodromy is reduced, is defined by
 $\{a, b, c_1-a, c_1-b, c_2-a, c_2-b, c_1 +c_2-a, c_1 +c_2-b\} \subset \mathbb{Z}$
 \cite{Bytev:2011ks}, and, in the case of Eq.~(\ref{bubble:eq:th}), we have
 $-b+c_1+c_2=1$, so that one solution of the
 PDE system (\ref{eq_F4}) degenerates to the Puiseux type, and the one-variable
 differential equation for $F_4$, Eq.~(\ref{eq:ex:4}), must factorize in such a
 way that a first-order differential operator splits off,
\begin{gather}
	L_1(x)L_3(x)J(\alpha_1,\alpha_2,m_1,m_2)=0 \,.
\end{gather}
 In terms of hypergeometric functions, the answer may be written as four independent solutions: three $F_4$ functions with various arguments and one polynomial. By defining suitable constants, we may
find that the final answer for the one-loop two-point FI with two different masses and arbitrary powers of propagators has only two $F_4$ terms, in the variables
$z_1=p^2/m_2^2$, $z_2=m_1^2/m_2^2$, and three terms in the variables 
$z_1=m_1^2/p^2$, $z_2=m_2^2/p^2$ \cite{Boos:1990rg}.

Let us now consider the case when the continuous variables $\vec{z}$ take the
same values, $z_1=z_2$.
In this case, $J$ is equivalent to the FI (\ref{bubble_FI}) with equal masses
$m_1=m_2$, and we have $z_1=y_1(x)=x$ and $z_2=y_2(x)=x$. 
As this univariate specialization does not belong to the singular locus of
Eqs.~(\ref{bubble:eq:th}) or (\ref{eq_F4}), the rank of the new PDE system is
the same as that of the original one.
Indeed, for the PDE system related to the Appell hypergeometric function $F_4$,
we obtain the following ordinary differential equation of fourth order in one
variable:
\begin{gather}
\tilde{L}_4(x)F_4(x,x)=0\,,    
\end{gather}
which has three distinct poles at points $0,1/4,\infty$.
Comparing the singular points and local exponents with the differential
equation for the hypergeometric function ${}_4F_3$, we recover the well-known
result for the univariate specialization of $F_4$
\cite{10.1093/qmath/os-13.1.90}:
\begin{gather}
F_4 \left( \begin{array}{c|}
a,b \\
c_1,c_2
\end{array}~ x,x \right)
=
{_4}F_3 \left( \begin{array}{c|}
a,b,\frac{c_1+c_2}{2},\frac{c_1+c_2-1}{2} \\
c_1,c_2,c_1+c_2-1
\end{array}~ 4x \right)\,.
\end{gather}

Above, we found that the monodromy group of the initial PDE system for the
considered FI in Eq.~(\ref{bubble:eq:th}) is reduced due to the constraint
$-b+c_1+c_2=1$ on its parameters.
We thus find the factorization of $\tilde{L}_4(x)$ by substituting the
parameters $a,b,c_1,c_2$ from Eq.~(\ref{bubble:eq:th}):
\begin{gather}
	L_1(x)L_3(x)J(\alpha_1,\alpha_2,m,m)=0\,,
\nonumber \\
 	 L_1(x)= \frac{\d}{\d x}+ \frac{ (x-4) (- {\alpha_1}- {\alpha_2}+d)+3 x-8}{ x(x-4)}\,,
 \nonumber	\\
 L_3(x)=
 \frac{\d^3}{\d x^3}+
 \frac{ -(x-8) ( {\alpha_1}+ {\alpha_2}-d-3)+2 d+18}{(x-4) x}\,
  \frac{\d^2}{\d x^2}
 \nonumber \\ 
 -\frac{ 4 \left(( {\alpha_1}+ {\alpha_2}) (5 ( {\alpha_1}+ {\alpha_2})-8 d+1)+3 d^2\right)+x (2  {\alpha_1}-d-2) (-2  {\alpha_2}+d+2)}{4 (x-4) x^2}\,
  \frac{\d}{\d x}
\nonumber  \\ 
 +\frac{( {\alpha_1}+ {\alpha_2}-d+1) ( {\alpha_1}+ {\alpha_2}-d+2) (2 ( {\alpha_1}+ {\alpha_2})-d)}{2 (x-4) x^3}\,.
 \end{gather}
As a consequence, the final answer for the one-loop two-point FI with equal masses can
be expressed through the hypergeometric function $_3F_2$ and a polynomial
expression, which may be found in Eqs.~(17) and (18) of
Ref.~\cite{Boos:1990rg}.

\section{Conclusions}
\label{conclusion}

In this work, we proposed a systematic method for deriving a system of PDEs for
a FI whose initial set of Lorentz invariants and masses may be arbitrarily
constrained, down to one or more free parameters.
This method does not rely on IBP relations and is applicable also for
non-integer propagator indices.
It proceeds in two steps.
In the first step, we treat all external momenta and masses as independent and
derive a prototype system of PDEs from the Mellin--Barnes representation of the
FI.
In the second step, we implement the constraints among the external momenta and
masses through a multivariate specialization and construct a new system of
PDEs for the particular FI.
This method also enables one to conveniently determine, during the second
step, the rank of the final PDE system, the number of its rational solutions,
and the simplest class of special functions through which the particular FI may
be expressed.

\section*{Acknowledgements}

We are grateful to A.~I.~Onishchenko and M.~Yu.~Kalmykov for fruitful
discussions.
The work of V.V.B. was supported in part by the Heisenberg--Landau Program.
The work of B.A.K. and O.L.V. was supported in part by the German Research
Foundation DFG through Research Unit FOR~2926 ``Next Generation Perturbative
QCD for Hadron Structure: Preparing for the Electron-Ion Collider" with Grant
Nos.~KN~365/13-1 and KN~365/14-1.

\bibliographystyle{plain}
\bibliography{references} 

\begin{thebibliography}{10}

\bibitem{Kotikov:1990kg}
A.~V. Kotikov.
\newblock {Differential equations method. New technique for massive Feynman
  diagram calculation}.
\newblock {\em Phys. Lett. B}, 254:158--164, 1991.

\bibitem{Kotikov:1991hm}
A.~V. Kotikov.
\newblock {Differential equations method: the calculation of vertex-type
  Feynman diagrams}.
\newblock {\em Phys. Lett. B}, 259:314--322, 1991.

\bibitem{Kotikov:1991pm}
A.~V. Kotikov.
\newblock {Differential equation method. The calculation of $N$-point Feynman
  diagrams}.
\newblock {\em Phys. Lett. B}, 267:123--127, 1991.
\newblock [Erratum: Phys. Lett. B 295, 409 (1992)].

\bibitem{Kotikov:1990zs}
A.~V. Kotikov.
\newblock {New method of massive Feynman diagrams calculation}.
\newblock {\em Mod. Phys. Lett. A}, 6:677--692, 1991.

\bibitem{Remiddi:1997ny}
E.~Remiddi.
\newblock {Differential equations for Feynman graph amplitudes}.
\newblock {\em Nuovo Cim. A}, 110:1435--1452, 1997.

\bibitem{Chetyrkin:1981qh}
K.~G. Chetyrkin and F.~V. Tkachov.
\newblock {Integration by parts: the algorithm to calculate $\beta$-functions
  in 4 loops}.
\newblock {\em Nucl. Phys. B}, 192:159--204, 1981.

\bibitem{Boos:1990rg}
\'E.~\'E. Boos and A.~I. Davydychev.
\newblock {A method of calculating massive Feynman integrals}.
\newblock {\em Theor. Math. Phys.}, 89:1052--1064, 1991.

\bibitem{Smirnov:2004ym}
V.~A. Smirnov.
\newblock {Evaluating Feynman Integrals}.
\newblock {\em Springer Tracts Mod. Phys.}, 211:1--244, 2004.

\bibitem{tHooft:1972tcz}
G.~'t~Hooft and M.~Veltman.
\newblock {Regularization and renormalization of gauge fields}.
\newblock {\em Nucl. Phys. B}, 44:189--213, 1972.

\bibitem{Kalmykov:2008ofy}
M.~Kalmykov, V.~Bytev, B.~Kniehl, B.~F.~L. Ward, and S.~A. Yost.
\newblock {Feynman Diagrams, Differential Reduction and Hypergeometric
  Functions}.
\newblock {\em PoS}, ACAT08:125, 2009.

\bibitem{Kalmykov:2012rr}
M.~Yu. Kalmykov and B.~A. Kniehl.
\newblock {Mellin-Barnes representations of Feynman diagrams, linear systems of
  differential equations, and polynomial solutions}.
\newblock {\em Phys. Lett. B}, 714:103--109, 2012.

\bibitem{Kalmykov:2020cqz}
M.~Kalmykov, V.~Bytev, B.~A. Kniehl, S.-O. Moch, B.~F.~L. Ward, and S.~A. Yost.
\newblock {Hypergeometric Functions and Feynman Diagrams}.
\newblock In {\em {Anti-Differentiation and the Calculation of Feynman
  Amplitudes}}, pages 189--234, 2021.

\bibitem{Bytev:2009kb}
V.~V. Bytev, M.~Yu. Kalmykov, and B.~A. Kniehl.
\newblock {Differential reduction of generalized hypergeometric functions from
  Feynman diagrams: One-variable case}.
\newblock {\em Nucl. Phys. B}, 836:129--170, 2010.

\bibitem{Bytev:2011ks}
V.~V. Bytev, M.~Yu. Kalmykov, and B.~A. Kniehl.
\newblock {HYPERDIRE, HYPERgeometric functions DIfferential REduction:
  MATHEMATICA-based packages for differential reduction of generalized
  hypergeometric functions ${}_pF_{p-1}$, $F_1$, $F_2$, $F_3$, $F_4$}.
\newblock {\em Comput. Phys. Commun.}, 184:2332--2342, 2013.

\bibitem{Bytev:2013gva}
V.~V. Bytev, M.~Yu. Kalmykov, and S.-O. Moch.
\newblock {HYPERgeometric functions DIfferential REduction (HYPERDIRE):
  MATHEMATICA based packages for differential reduction of generalized
  hypergeometric functions: $F_D$ and $F_S$ Horn-type hypergeometric functions
  of three variables}.
\newblock {\em Comput. Phys. Commun.}, 185:3041--3058, 2014.

\bibitem{Bytev:2013bqa}
V.~V. Bytev and B.~A. Kniehl.
\newblock {HYPERDIRE HYPERgeometric functions DIfferential REduction:
  Mathematica-based packages for the differential reduction of generalized
  hypergeometric functions: Horn-type hypergeometric functions of two
  variables}.
\newblock {\em Comput. Phys. Commun.}, 189:128--154, 2015.

\bibitem{Bytev:2016ibi}
V.~V. Bytev and B.~A. Kniehl.
\newblock {HYPERDIRE\textemdash{}HYPERgeometric functions DIfferential
  REduction: Mathematica-based packages for the differential reduction of
  generalized hypergeometric functions: Lauricella function $F_C$ of three
  variables}.
\newblock {\em Comput. Phys. Commun.}, 206:78--83, 2016.

\bibitem{Kalnins1980TheLT}
E.~G. Kalnins, H.~L. Manocha, and W.~Miller, Jr.
\newblock The lie theory of two-variable hypergeometric functions.
\newblock {\em Studies in Applied Mathematics}, 62:143--173, 1980.

\bibitem{VIDUNAS2009145}
R.~Vid{\=u}nas.
\newblock Specialization of appell's functions to univariate hypergeometric
  functions.
\newblock {\em J. Math. Anal. Appl.}, 355:145--163, 2009.

\bibitem{10.1093/qmath/os-13.1.90}
J.~L. Burchall.
\newblock {Differential equations associated with hypergeometric functions}.
\newblock {\em The Quarterly Journal of Mathematics}, os-13(1):90--106, 01
  1942.

\end{thebibliography}
\end{document}